\documentstyle[psfig]{mn}

\newif\ifAMStwofonts

\def\H0{{$H_0$}}

\def\Sub#1{_{_{#1}}}

\def\Mod#1{\mu\Sub{#1}}

\def\imm#1{\raise0.4pt\hbox{$\langle$}$#1$\raise0.4pt\hbox{$\rangle$}}
\def\Imm#1{{\raise0.4pt\hbox{$\langle$}{#1}\raise0.4pt\hbox{$\rangle$}}}
\def\Square#1{{\raise0.4pt\hbox{$[$}{#1}\raise0.4pt\hbox{$]$}}}

\def\lsim{{\small\mathrel{\hbox{\rlap{\hbox{\lower2pt\hbox{$\sim$}}}\raise2pt\hbox{$<$}}}}}
\def\gsim{{\small\mathrel{\hbox{\rlap{\hbox{\lower2pt\hbox{$\sim$}}}\raise2pt\hbox{$>$}}}}}
\def\ie{{\it i.e.}}
\def\eg{{\it e.g.}}

\def\etal{{\it et~al.}}

\ifoldfss
  \ifCUPmtlplainloaded \else
    \NewTextAlphabet{textbfit} {cmbxti10} {}
    \NewTextAlphabet{textbfss} {cmssbx10} {}
    \NewMathAlphabet{mathbfit} {cmbxti10} {} 
    \NewMathAlphabet{mathbfss} {cmssbx10} {} 
  \fi
  \ifAMStwofonts
    \ifCUPmtlplainloaded \else
      \NewSymbolFont{upmath} {eurm10}
      \NewSymbolFont{AMSa} {msam10}
      \NewMathSymbol{\upi}     {0}{upmath}{19}
      \NewMathSymbol{\umu}     {0}{upmath}{16}
      \NewMathSymbol{\upartial}{0}{upmath}{40}
      \NewMathSymbol{\leqslant}{3}{AMSa}{36}
      \NewMathSymbol{\geqslant}{3}{AMSa}{3E}

    \fi
  \fi
\fi 

\ifnfssone
  \newmathalphabet{\mathit}
  \addtoversion{normal}{\mathit}{cmr}{m}{it}
  \addtoversion{bold}{\mathit}{cmr}{bx}{it}
  \newmathalphabet{\mathbfit} 
  \addtoversion{normal}{\mathbfit}{cmr}{bx}{it}
  \addtoversion{bold}{\mathbfit}{cmr}{bx}{it}
  \newmathalphabet{\mathbfss} 
  \addtoversion{normal}{\mathbfss}{cmss}{bx}{n}
  \addtoversion{bold}{\mathbfss}{cmss}{bx}{n}
  \ifAMStwofonts
    \ifCUPmtlplainloaded \else
      \UseAMStwoboldmath
      \makeatletter
      \new@mathgroup\upmath@group
      \define@mathgroup\mv@normal\upmath@group{eur}{m}{n}
      \define@mathgroup\mv@bold\upmath@group{eur}{b}{n}
      \edef\UPM{\hexnumber\upmath@group}
      \new@mathgroup\amsa@group
      \define@mathgroup\mv@normal\amsa@group{msa}{m}{n}
      \define@mathgroup\mv@bold\amsa@group{msa}{m}{n}
      \edef\AMSa{\hexnumber\amsa@group}
      \makeatother
      \mathchardef\upi="0\UPM19
      \mathchardef\umu="0\UPM16
      \mathchardef\upartial="0\UPM40
      \mathchardef\leqslant="3\AMSa36
      \mathchardef\geqslant="3\AMSa3E
    \fi
  \fi
\fi 

\ifnfsstwo
  \DeclareMathAlphabet{\mathbfit}{OT1}{cmr}{bx}{it}
  \SetMathAlphabet\mathbfit{bold}{OT1}{cmr}{bx}{it}
  \DeclareMathAlphabet{\mathbfss}{OT1}{cmss}{bx}{n}
  \SetMathAlphabet\mathbfss{bold}{OT1}{cmss}{bx}{n}
  \ifAMStwofonts
    \ifCUPmtlplainloaded \else
      \DeclareSymbolFont{UPM}{U}{eur}{m}{n}
      \SetSymbolFont{UPM}{bold}{U}{eur}{b}{n}
      \DeclareSymbolFont{AMSa}{U}{msa}{m}{n}
      \DeclareMathSymbol{\upi}{0}{UPM}{"19}
      \DeclareMathSymbol{\umu}{0}{UPM}{"16}
      \DeclareMathSymbol{\upartial}{0}{UPM}{"40}
      \DeclareMathSymbol{\leqslant}{3}{AMSa}{"36}
      \DeclareMathSymbol{\geqslant}{3}{AMSa}{"3E}
    \fi
  \fi
\fi 

\ifCUPmtlplainloaded \else
  \ifAMStwofonts \else 
    \def\upi{\pi}
    \def\umu{\mu}
    \def\upartial{\partial}
  \fi
\fi

\title{Photometry of 40 LMC Cepheids}
\author[N. R. Tanvir and A. Boyle]
       {N. R. Tanvir$^{1,2}$ and A. Boyle$^3$ \\
       $^1$Institute of Astronomy, University of Cambridge, 
       Madingley Road, Cambridge, CB3 0HA. UK.\\
       $^2$Department of Physical Sciences, University of Hertfordshire,
       College Lane, Hatfield, Herts. AL10 9AB. UK.\\
       $^3$Department of Physics, National University of Ireland,
       Galway, Ireland.}

\date{Accepted .
      Received ;
      in original form }

\pagerange{\pageref{firstpage}--\pageref{lastpage}}

\begin{document}

\maketitle

\label{firstpage}

\begin{abstract}
We present $V$ and $I_c$ CCD photometry for 40 LMC
Cepheids at 1 to 3 epochs.  This represents a significant increase in the
number of LMC Cepheids with $I$-band data, and, as we show, is a useful 
addition to the sample which can be used to calibrate the 
period--luminosity relations in these important bands.
\end{abstract}


\section{Introduction}

The bulk of recent extra-galactic Cepheid studies have used $V$-band
observations to search for variability and characterise the light
curves, and $I$-band observations to give colours and hence allow
a correction for reddening (\eg\ Tanvir \etal\ 1995).
Usually this involves calculating apparent distance moduli
in both bands and calculating the true modulus $\Mod{0}=\Mod{AV}-R\Sub{VI}(\Mod{AV}-\Mod{AI})$.
As emphasized by Tanvir (1997; hereafter T97), this approach is 
equivalent to determining
reddening corrected Wesenheit indices for the Cepheids
(see also Madore 1982; van den Bergh 1968), defined as

$$W\Sub{VI}=\Imm{V}-R\Sub{VI}\Square{\Imm{V}-\Imm{I}}$$

with $R\Sub{VI}=A\Sub{V}/E\Sub{V-I}$,
and fitting a suitable PL relation to them.
However, the limited amount of photoelectric
$I$-band data for Cepheids in the LMC is an impediment to calibrating this
PL relation.

Fortunately reasonably good Wesenheit indices can be determined from
observations at relatively few epochs.
This is because the natural
variations in colour and luminosity around a pulsation cycle mimic
the effects of dust \ie\ at their brightest the Cepheids are
also at their bluest (see Madore 1985 for discussion in context
of the ``Feinheit'' method).
We illustrate this in figure 1 where we have taken the densely-sampled,
high-quality data from Moffett \etal\ (1998) 
for several high-amplitude Cepheids, and resampled it many (10000) times
at two randomly chosen epochs to see how the calculated value of
$W$ compares with that found from the full data-sets.
The {\it rms} dispersion of the estimates around the true value
is only 0.13 mags.

\begin{figure}
 \centerline{\psfig{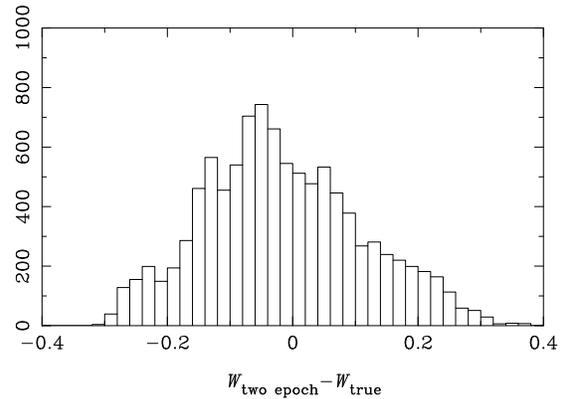}}
 \caption{Histogram showing the error made in determining
a Cepheid's Wesenheit index from observations at two, randomly
chosen epochs.  This is based on resampling many times the
dense, high-quality data for LMC Cepheids from Moffett \etal\ (1998).
These Cepheids have periods in the region of 30 days and so are
large amplitude and hence worst case.  The formal {\it rms} dispersion is
only 0.13 magnitudes, but note that there is a small offset in the mean
of -0.016 mags which is a consequence of following the traditional approach
of calculating magnitudes at mean intensity.
}
\end{figure}

\medskip

Here we report CCD observations of
a large number (40) of LMC Cepheids, whose periods are 
already known from photographic work,
at 1 to 3 epochs over 6 nights.
The data presented will be combined with other data from
the literature in a future publication to determine new PL relations
(Tanvir, in prep.).

\section{Observations and analysis}

Our observations were obtained on the nights of the $14^{\rm th}$
to  $19^{\rm th}$ of November 1996 with the Danish 1.5 m telescope
at La Silla.
The DFOSC camera was equipped with a $2048\times2048$ pixel,
thinned Loral CCD (W11-4),
which with 0.39 arcsec pixels gave a 13.3 arcmin field of view.
Unfortunately this chip was cosmetically poor around the edges
and so we restricted the analysis to a circular region around
the centre of radius 800 pixels.

The 28 primary targets were chosen to be Cepheids with periods
between 8 and 50 days, the range explored in most HST extragalactic
studies, which have
little or no previous photoelectric $I$-band
photometry, but often some $V$-band photometry.
Finding charts from Hodge \& Wright (1967) were used to locate
the variables on the frames.
The remainder of the sample consists of other Cepheids, usually
of shorter period, which happened to lie in the same fields
and which typically have no other photoelectric photometry.

$V$- and $I$-band  exposures were obtained at
each epoch, with exposure times ranging from 10 s to 60 s.
Every night was photometric, and we obtained flat-fields and
multiple standard star observations (specifically fields in
SA95, SA98, SA114 and around T Phe from Landolt 1992) so that each 
night could be calibrated independently onto the $VI_c$ systems.
In practice the zero-points of the magnitude scales agreed from
night to night to  0.01 mag.
Colour terms were determined by combining all the standard star
photometry and, for the difference in the average colour between
the standards and the Cepheids, amounted to less than 0.01 mags
in each case.
Although the seeing varied, sometimes quite rapidly, between
about 0.9 arcsec and about 1.7 arcsec, we found that magnitudes measured
in a 6 arcsec 
diameter aperture, over this range,
were not very sensitive to the seeing.

Nightly extinction coefficients were taken from the
data-base of the Geneva Observatory Photometric group 
(http://obswww.unige.ch/photom/extlast.html; see Burki \etal\ 1995),
and range between 0.12 and 0.14 mag per air-mass for the $V$-band, in
excellent agreement
with our standard star observations.
Although $I$-band  extinction coefficients were not available for
the nights of our run, we adopted a value of 0.06 mag per airmass
based on the typical values for other nights which were tabulated.
Air-masses for the observations were typically in the range 1.3 to 1.6,
which is inevitable given the declination of the LMC, while the standard
fields, although observed at a wide range of air-masses,  were in most cases
lower at 1.1 to 1.2.

Each frame was debiassed and flat-fielded in the normal way.
Interactive aperture photometry was performed with the apphot.phot 
routine within 
IRAF, for the target Cepheids, standard stars and also for several field
stars in each frame.
Each star was measured in apertures of 2, 4, 8, 16 and 32 pixels radius.
In most cases, the 4 pixel radius aperture was used, to minimize any small
crowding errors, and an aperture correction to 16 pixels 
was determined from a number of stars
in the frame and other frames of similar seeing taken on the same
night.
The sky level was determined from the pixels in a large annulus around the
program star.

\section {Results}

The photometry is listed in table 1.  For each variable the
period is given in parentheses and  the three columns are
(1) the modified Julian date of observations, (2) $V$ and (3) $I_c$.
Since each night was calibrated independently, it is possible to get
fairly good estimates of the true photometric errors by comparing
the magnitudes of non-variable stars observed on different nights.
In figure 2 this is done for a set of stars which were observed
on 3 occasions, and shows that typical  errors in the 
magnitude range of interest are
around 0.015 magnitudes.  Although small, this is greater than the
formal errors reported by phot, showing that, as expected, 
the calibration and
aperture corrections are also important sources of uncertainty.
This also explains why the dispersion increases very little with
magnitude.
As another test of our photometry, we also observed the
two LMC photometric standard stars CPD66349 and CPD66350 
(Menzies \etal\ 1989).
First transforming the standard magnitudes to Landolt's (1983) system
via the equations given in Menzies \etal\ (1991) we obtain 
the following differences in the sense of {\it us} minus {\it standard}
for the two stars: -0.022 and -0.003 in $V$ and -0.017 and -0.017 in $I_c$.
Again this is reasonably consistent with a typical error of 0.015 mags.

\begin{figure}
\centerline{\psfig{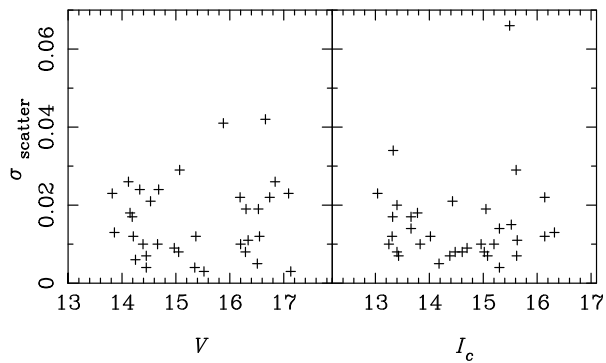}}
 \caption{Estimates of the photometric error from the 
scatter in the magnitudes of
(assumed) non-variable stars observed on three occasions. 
Since each night is calibrated independently, this gives an
indication of the true photometric errors.
Although
there may be a small increase in 
$\sigma_{\rm scatter}$  at faint magnitudes, at
bright magnitudes it is certainly dominated by the calibration uncertainties
and the limiting value is around 0.015 mags in each case.
}
\end{figure}

\begin{table*}
 \centering
 \begin{minipage}{120mm}
  \caption{$V$ and $I_c$ photometry for our sample.  The periods in days, 
given in parentheses, are either taken from Payne-Gaposchkin (1971), or are 
re-determined by us for those cases where other photoelectric 
$V$-band photometry exists in the literature.
}
  \begin{tabular}{@{}rrrrrrrrr@{}}
\multicolumn{3}{c}{{\it HV873}\quad(34.427)} & \multicolumn{3}{c}{{\it HV875}\quad(30.328)} & \multicolumn{3}{c}{{\it HV878}\quad(23.304)} \\
50402.32 & 13.406 & 12.256    & 50402.32 & 13.286 & 12.326    & 50402.32 & 13.508 & 12.570 \\
50407.06 & 13.636 & 12.524    & 50407.07 & 13.100 & 12.268    & 50402.33 & 13.507 & 12.546 \\
&& & &&                                                       & 50407.07 & 13.845 & 12.798 \\
\multicolumn{3}{c}{{\it HV881}\quad(35.743)} & \multicolumn{3}{c}{{\it HV882}\quad(31.795)} & \multicolumn{3}{c}{{\it HV886}\quad(23.98)} \\
50402.33 & 12.561 & 11.900    & 50402.33 & 12.793 & 12.092    & 50403.33 & 13.755 & 12.771  \\
50407.07 & 12.754 & 11.932    & 50407.08 & 13.008 & 12.167    & 50407.08 & 13.193 & 12.496  \\
\multicolumn{3}{c}{{\it HV889}\quad(25.805)} & \multicolumn{3}{c}{{\it HV900}\quad(47.51)} & \multicolumn{3}{c}{{\it HV902}\quad(26.346)} \\
50403.33 & 14.051 & 12.906    & 50403.34 & 13.026 & 11.912    & 50402.36 & 12.796 & 12.125  \\
50407.09 & 14.038 & 12.986    & &&                            & 50407.09 & 13.065 & 12.213  \\
\multicolumn{3}{c}{{\it HV909}\quad(37.565)} & \multicolumn{3}{c}{{\it HV955}\quad(13.737)} & \multicolumn{3}{c}{{\it HV1003}\quad(24.345)} \\
50402.36 & 12.865 & 11.881    & 50402.35 & 13.526 & 12.926    & 50403.35 & 13.550 & 12.561  \\
50407.10 & 12.974 & 11.995    & 50406.22 & 14.035 & 13.162    & 50407.12 & 13.549 & 12.631  \\
&&                            & 50407.11 & 14.120 & 13.208    & &&                          \\
\multicolumn{3}{c}{{\it HV1005}\quad(18.711)} & \multicolumn{3}{c}{{\it HV2251}\quad(27.916)} & \multicolumn{3}{c}{{\it HV2254}\quad(3.168)} \\
50403.35 & 13.702 & 12.971    & 50403.33 & 13.539 & 12.454    & 50403.33 & 16.198 & 15.459  \\
50407.13 & 13.991 & 13.006    & 50407.08 & 13.639 & 12.610    & 50407.08 & 15.370 & 14.924  \\
\multicolumn{3}{c}{{\it HV2257}\quad(39.37)} & \multicolumn{3}{c}{{\it HV2291}\quad(22.328)} & \multicolumn{3}{c}{{\it HV2295}\quad(7.846)} \\
50402.32 & 13.113 & 11.982    & 50403.33 & 14.335 & 13.301    & 50403.33 & 14.855 & 14.038  \\
50402.33 & 13.123 & 11.995    & 50407.09 & 13.640 & 12.791    & 50407.09 & 15.201 & 14.292  \\
50407.07 & 13.255 & 12.122    & &&                            & &&                          \\
\multicolumn{3}{c}{{\it HV2432}\quad(10.925)}  & \multicolumn{3}{c}{{\it HV2523}\quad(6.784)}   & \multicolumn{3}{c}{{\it HV2527}\quad(12.949)} \\
50402.34 & 14.585 & 13.690    & 50402.35 & 15.089 & 14.207    & 50402.35 & 14.209 & 13.50   \\
50404.37 & 14.329 & 13.607    & 50405.36 & 14.825 & 14.061    & 50405.36 & 14.379 & 13.52   \\
&&                            & 50407.11 & 14.987 & 14.148    & 50407.11 & 14.550 & 13.579  \\
\multicolumn{3}{c}{{\it HV2549}\quad(16.216)} & \multicolumn{3}{c}{{\it HV2579}\quad(13.425)} & \multicolumn{3}{c}{{\it HV2662}\quad(12.075)} \\
50402.36 & 14.097 & 13.191    & 50406.23 & 14.456 & 13.490    & 50403.34 & 14.546 & 13.533  \\
50407.11 & 13.146 & 12.624    & 50407.12 & 14.314 & 13.430    & 50405.37 & 14.631 & 13.655  \\
&& & &&                                                       & 50407.12 & 14.406 & 13.544  \\
\multicolumn{3}{c}{{\it HV2722}\quad(8.027)} & \multicolumn{3}{c}{{\it HV2738}\quad(8.337)} & \multicolumn{3}{c}{{\it HV5511}\quad(3.340)} \\
50403.36 & 14.941 & 14.134    & 50403.36 & 14.418 & 13.760    & 50402.32 & 16.299 & 15.431  \\
50404.36 & 14.561 & 13.911    & 50404.36 & 14.496 & 13.759    & 50407.06 & 15.742 & 15.096  \\
50407.13 & 14.386 & 13.742    & 50407.13 & 14.924 & 14.023    & &&                          \\
\multicolumn{3}{c}{{\it HV6105}\quad(10.440)} & \multicolumn{3}{c}{{\it HV8036}\quad(28.38)} & \multicolumn{3}{c}{{\it HV12248}\quad(10.912)} \\
50402.34 & 15.104 & 14.183    & 50402.30 & 13.863 & 12.705    & 50403.36 & 14.763 & 13.800  \\
50404.36 & 14.617 & 13.892    & 50407.04 & 14.001 & 12.864    & 50406.23 & 14.475 & 13.692  \\
50407.10 & 14.844 & 13.937    & &&                            & 50407.14 & 14.361 & 13.620  \\
\multicolumn{3}{c}{{\it HV12253}\quad(12.574)} & \multicolumn{3}{c}{{\it HV12416}\quad(3.928)} & \multicolumn{3}{c}{{\it HV12426}\quad(2.550)} \\
50404.35 & 14.790 & 13.857    & 50402.32 & 15.994 & 15.088    & 50402.33 & 15.795 & 15.264  \\
50406.24 & 14.644 & 13.771    & 50407.06 & 16.087 & 15.245    & 50407.07 & 15.515 & 15.150  \\
50407.14 & 13.779 & 13.256    & &&                          & &&                        \\
\multicolumn{3}{c}{{\it HV12471}\quad(15.851)} & \multicolumn{3}{c}{{\it HV12503}\quad(2.731)} & \multicolumn{3}{c}{{\it HV12619}\quad(3.481) 
\footnote{Note that there is some 
uncertainty about the period  of HV12619, which is given 
by Payne-Gaposchkin (1971) 
as 2.480646 , but whose position within her table 5 suggests a typographical 
error and that the leading number should be a 3.  However, 
given that, in addition, this variable 
is  flagged as having significant scatter, we recommend it be treated 
with caution.}} \\
50402.31 & 14.857 & 13.757    & 50402.32 & 15.861 & 15.277    & 50403.34 & 15.295 & 14.578  \\
50405.36 & 14.309 & 13.429    & 50407.07 & 16.260 & 15.581    & 50405.37 & 15.215 & 14.535  \\
50407.05 & 14.432 & 13.448    & &&                            & 50407.12 & 15.230 & 14.576  \\
\multicolumn{3}{c}{{\it HV12716}\quad(11.248)} & \multicolumn{3}{c}{{\it HV12787}\quad(3.676)} & \multicolumn{3}{c}{{\it U1}\quad(22.54)}      \\
50402.30 & 14.391 & 13.535    & 50402.34 & 15.375 & 14.847    & 50402.31 & 14.594 & 13.325  \\
50405.36 & 14.837 & 13.800    & 50404.36 & 15.955 & 15.170    & 50407.04 & 14.570 & 13.447  \\
50407.03 & 15.020 & 13.960    & 50407.10 & 15.667 & 14.973    &  &&                         \\
\multicolumn{3}{c}{{\it U11}\quad(20.077)}     & &&  & &&                        \\
50402.31 & 14.337 & 13.201    & &&                            & && \\
50407.05 & 14.249 & 13.275    & &&                            & && \\

\end{tabular}
\end{minipage}
\end{table*}

\section {Discussion}

As we have said, the primary motivation for obtaining this data is
to combine it with other data from the literature to provide
a large sample of Cepheids with which to
explore the calibration of the Cepheid period--luminosity
relations in the LMC (Tanvir in prep.).
Here we simply plot the intensity mean magnitudes in each
band and Wesenheit indices (figure 3) where we have taken $R\Sub{VI}=2.45$
(T97).
This demonstrates
that, even with a small number of epochs,
the Wesenheit indices  indeed produce an
impressively tight PL relation.

\begin{figure}
\centerline{\psfig{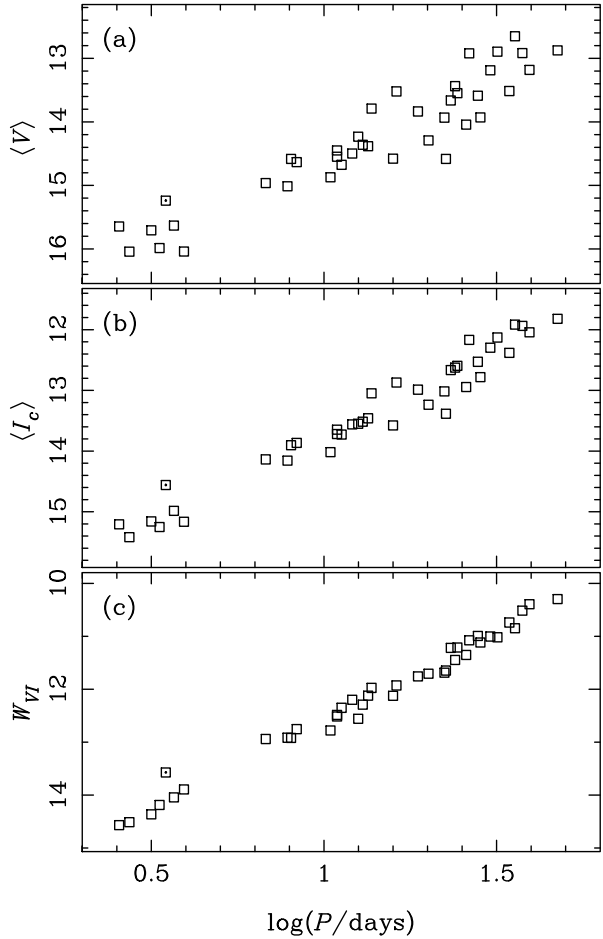}}
 \caption{Intensity mean magnitudes in (a) $V$ and (b) $I_c$,
and (c) Wesenheit indices for our sample plotted against log period.
We have not corrected here for the small bias in the intensity
means caused by the undersampling, which was pointed out in the caption
to figure 1.
Note that the scatter is
large in $V$ and $I_c$, but much smaller in $W\Sub{VI}$ as expected.
The most extreme outlier, indicated by a central dot, is in fact HV12619 
which, as noted in
the caption to table 1, is of uncertain status and period.
}
\end{figure}

\section*{Acknowledgments}

AB acknowledges an RGO summer studentship.

\end{document}